\documentclass[12pt]{article}
\usepackage{graphicx}
\usepackage{amsmath}
\usepackage{longtable}

\begin{document}

\makeatletter
\renewcommand*{\@cite}[2]{{#2}}
\renewcommand*{\@biblabel}[1]{#1.\hfill}
\makeatother

\title{Influence of the Gould Belt on Interstellar Extinction}
\author{G.~A.~Gontcharov\thanks{E-mail: georgegontcharov@yahoo.com}}

\maketitle

Pulkovo Astronomical Observatory, Russian Academy of Sciences, Pul\-kov\-skoe sh. 65, St. Petersburg, 196140 Russia


A new analytical 3D model of interstellar extinction within 500 pc of the Sun as a function
of the Galactic spherical coordinates is suggested. This model is physically more justified than the widely
used Arenou model, since it takes into account the presence of absorbing matter both in the layer along
the equatorial Galactic plane and in the Gould Belt. The extinction in the equatorial layer varies as the
sine of the Galactic longitude and in the Gould Belt as the sine of twice the longitude in the Belt plane.
The extinction across the layers varies according to a barometric law. It has been found that the absorbing
layers intersect at an angle of $17^{\circ}$ and that the Sun is located near the axial plane of the absorbing layer of
the Gould Belt and is probably several parsecs below the axial plane of the equatorial absorbing layer but
above the Galactic plane. The model has been tested using the extinction of real stars from three catalogs.

\newpage
\section*{INTRODUCTION}

There are regions with a comparatively high interstellar extinction far from the Galactic plane. For
example, it can be seen on the $(J-Ks)$ -- $Ks$ diagrams
for stars from the 2MASS catalog (Skrutskie et al. 2006) with accurate photometry in the $J$ and $Ks$ infrared
bands presented in Fig. 1 that the reddening
of stars in the regions at $l\approx180^{\circ}$, $b\approx-15^{\circ}$ and
$l\approx0^{\circ}$, $b\approx+15^{\circ}$ is considerably higher than that in the
symmetric regions at $l\approx180^{\circ}$, $b\approx+15^{\circ}$ and
$l\approx0^{\circ}$, $b\approx-15^{\circ}$. Since the red dwarfs with $(J-Ks)\approx0.9^{m}$,
$Ks\approx13^{m}$ and $M_{Ks}\approx5$, thus located at distances
up to 400 pc, reddened in these regions in the same
way as distant stars, it should be recognized that the
matter that caused this reddening is within 400 pc.

The model by Arenou et al. (1992), which approximates
the mean extinction $A_V$ for 199 regions of the
sky by parabolas depending on the distance is still
the best analytical 3D model of interstellar extinction
within the nearest kiloparsec. This model reproduces
adequately the observations within the nearest kiloparsec
of the Sun. Its drawback is the absence of
any physical explanation for the regularities in the
observed extinction variations.

Figure 2a shows the dependence of extinction
$A_V$ on the Galactic longitude calculated using the
model by Arenou et al. (1992) for a distance of 500 pc
and Galactic latitudes $+15^{\circ}<b<+30^{\circ}$, $+5^{\circ}<b<+15^{\circ}$, $-5^{\circ}<b<+5^{\circ}$,
$-15^{\circ}<b<-5^{\circ}$, $-30^{\circ}<b<-15^{\circ}$. The vertical bars indicate the
accuracy of the model. The general, approximately
sinusoidal longitude dependence of extinction near
the Galactic equator is shown in Fig. 2b as $0.8+0.5\sin(l+20^{\circ})$ (dashes). The Arenou model reflects
but does not explain this dependence as well as the
above-mentioned extinction features far from the
Galactic plane: we see from Fig. 2a that the extinction
is higher toward the Galactic center $+5^{\circ}<b<+30^{\circ}$
and toward the anticenter for $-30^{\circ}<b<-5^{\circ}$.

The accuracy of this model, i.e., the accuracy of
predicting the extinction for a specific star, is, on average,
from 20\% to 50\% at highGalactic latitudes and
about 40\% near the Galactic equator. This accuracy
reflects the actual extinction variations from star to
star in the same region of space: for example, for two
neighboring stars 500 pc away; a scatter in extinction
of $\pm0.3^{m}$ at a typical extinction $A_{V}=0.7^{m}$ is possible.
Thus,modeling the extinction within the nearest kiloparsec,
where each cloud, each Galactic structure,
and even each star plays a role, is much more difficult
than its modeling far from the Sun, where the high
mean extinction smoothes out the role of individual
objects, and the extinction for a typical large region
of space can be determined as, say, $5.0^{m}\pm0.5^{m}$, i.e.,
with an accuracy of 10\%. An \emph{analytical} model of
extinction within the nearest kiloparsec with a relative
accuracy higher than that of the model by Arenou
et al. (1992) is unlikely to be possible. To achieve a
high accuracy of the extinction correction, it is more
preferable to reveal specific absorbing clouds or to
determine the individual extinction for each star from
its highly accurate multiband photometry. However,
a physically justified analytical model is important for
analyzing large-scale structures within the nearest
kiloparsec.

A Galactic structure, the Gould Belt, that lies
outside the Galactic plane in the regions of the
sky where an enhanced extinction is observed and
that has a suitable size, several hundred parsecs,
exists in the neighborhood of the Solar system.
The Gould Belt and related Galactic structures, the
Local Bubble and the Great Tunnel, were described
by Gontcharov and Vityazev (2005 and references
therein), Bobylev (2006 and references therein),
and Perryman (2009, pp. 324-328 and references
therein). The Gould Belt contains young stars and
their associations. Stars are also formed here at
present. The accompanying interstellar clouds can
cause extinction. The extinction in the Gould Belt
was first pointed out by Vergely et al. (1998).

In this paper, we tested the hypothesis that the
interstellar extinction in the Gould Belt supplements
the extinction along the Galactic plane and contributes
significantly to the 3D pattern of extinction
within the nearest kiloparsec.

\section*{THE MODEL}

Figure 3 presents the relative positions of two
layers of absorbing matter, the layer with halfthickness
$Z_{A}$ near the equatorial Galactic plane
(below referred to as the equatorial layer) and the layer
with half-thickness $\zeta_{A}$ in the Gould Belt. Denote the
inclination of the Gould Belt to the Galactic plane
by $\gamma$. The working coordinate system is defined by
the observed coordinates of stars: the heliocentric
distance $r$ and the Galactic longitude $l$ and latitude $b$.
The Sun is at the origin of the working coordinate
system and we do not consider its displacement
relative to the Galactic plane, because it cannot
be determined in the model under consideration.
However, we consider the displacement of the axial
plane of the equatorial layer relative to the Sun, $Z_{0}$,
and the analogous displacement for the absorbing
layer of the Gould Belt, $\zeta_{0}$. For clarity and comparison
with the standard Galactic coordinate system, Fig. 3
shows the $X'$ and $Y'$ axes -- the axes of a rectangular
coordinate system in the axial plane of the equatorial
layer. The $X'$ axis is parallel to the direction toward
the Galactic center and the $Y'$ axis is parallel to the
direction of Galactic rotation. We will designate the
rotation of the highest point of the Gould Belt relative
to the $X'$ axis, i.e., the angle between the $Y'$ axis and
the line of intersection between the axial plane of the
equatorial layer and the axial plane of the Gould Belt,
as $\lambda_{0}$.

The longitude $\lambda$ and latitude $\beta$ of a star relative to
the axial plane of the Gould Belt can be calculated
from its Galactic coordinates:
\begin{equation}
\label{equ1}
\sin(\beta)=\cos(\gamma)\sin(b)-\sin(\gamma)\cos(b)\cos(l)
\end{equation}
\begin{equation}
\label{equ2}
\tan(\lambda-\lambda_{0})=\cos(b)\sin(l)/(\sin(\gamma)\sin(b)+\cos(\gamma)\cos(b)\cos(l)).
\end{equation}
The observed extinction $A$ is approximated by the
sum of two functions:
\begin{equation}
\label{aaa}
A=A(r,l,b)+A(r,\lambda,\beta),
\end{equation}
each of them is represented by a barometric law
(Parenago 1954, p. 265). The extinction in the equatorial
layer is
\begin{equation}
\label{aeq}
A(r,l,b)=(A_{0}+A_{1}\sin(l+A_{2}))Z_{A}(1-e^{-r|\sin(b)|/Z_{A}})/|\sin(b)|,
\end{equation}
where $A_0$, $A_1$, and $A_2$ are the free extinction term,
amplitude, and phase in the sinusoidal dependence
on $l$, and the extinction in the Gould Belt is
\begin{equation}
\label{ago}
A(r,\lambda,\beta)=(\Lambda_{0}+\Lambda_{1}\sin(2\lambda+\Lambda_{2}))\zeta_{A}(1-e^{-r|\sin(\beta)|/\zeta_{A}})/|\sin(\beta)|,
\end{equation}
where $\Lambda)0$, $\Lambda_1$, and $\Lambda_2$ are the free extinction term,
amplitude, and phase in the sinusoidal dependence
on $2\lambda$. The assumption that the extinction in the
Gould Belt has two maxima in the dependence on
longitude $\lambda$ was confirmed in our subsequent study.
The extinctionmaxima in the Gould Belt are observed
near the directions where the distance of the Belt from
the Galactic plane is at a maximum, i.e., approximately
in the directions of the Galactic center and
anticenter.

Given the displacement of the Sun relative to the
absorbing layers, Eqs. (4) and (5) transform to
\begin{equation}
\label{aeq2}
A(r,l,Z)=(A_{0}+A_{1}\sin(l+A_{2}))r(1-e^{-|Z-Z_{0}|/Z_{A}})Z_{A}/|Z-Z_{0}|
\end{equation}
and
\begin{equation}
\label{ago2}
A(r,\lambda,\zeta)=(\Lambda_{0}+\Lambda_{1}\sin(2\lambda+\Lambda_{2}))r(1-e^{-|\zeta-\zeta_{0}|/\zeta_{A}})\zeta_{A}/|\zeta-\zeta_{0}|.
\end{equation}
The quantities $|Z-Z_{0}|/Z_{A}$ and $|\zeta-\zeta_{0}|/\zeta_{A}$ that are
encountered in these formulas twice characterize the
stellar position in the absorbing layers displaced relative
to the Sun. We do not consider the displacement
of the Sun relative to the center of the Gould Belt,
because the accuracy of the data used is insufficient
for this purpose.

As a result, we have the system of equations (3),
one equation for each star. The observed extinction A
is on the left-hand sides and the function of three
observed quantities - $r$, $l$, and $b$ - is on the right-hand
sides. The solution gives 12 unknowns: $\gamma$, $\lambda_{0}$, $Z_{A}$, $\zeta_{A}$, $Z_{0}$, $\zeta_{0}$,
$A_{0}$, $A_{1}$, $A_{2}$, $\Lambda_{0}$, $\Lambda_{1}$, $\Lambda_{2}$. These are
chosen so as to minimize the sum of the squares of
the residuals of the left-hand and right-hand sides of
Eqs. (3).

We can estimate some of the unknowns in advance.
The inclination $\gamma$ of the Gould Belt to the
Galactic plane has been estimated by different researchers
to be in the range 10$^{\circ}$-25$^{\circ}$. Since themaximum
height of the Gould Belt above the Galactic
plane approximately coincides with the direction of
the Galactic center, $\lambda_{0}\approx0^{\circ}$. The half-thickness of
the absorbing layer $Z_{A}$ is close to 100 pc (Parenago
1954). If the absorbing layer of the Gould Belt
was produced by some ``deformation'' of the equatorial
layer, then we can also assume the same halfthickness
for it, i.e., $\zeta_{A}\approx100$ pc. The displacements
of the Sun relative to the absorbing layers $Z_{0}$ and $\zeta_{0}$ are unlikely to exceed several parsecs; otherwise,
this would be quite noticeable in the observational
data. The combined mean extinction in the equatorial
layer $A_{0}$ and the Gould Belt $\Lambda_{0}$ cannot differ too much
from the universally accepted extinction in the near
part of the Galaxy, $1.5^{m}$ kpc$^{-1}$.

\section*{COMPARISON OF THE MODELS}

Given the large scatter of individual extinctions for
stars within the nearest kiloparsec noted above, one
might expect better agrement of the suggested analytical
model and the model by Arenou et al. between
themselves than with observations. Therefore, comparing
the extinctions calculated using these models
for the same stars is of great importance. In this
case, the accuracy of the Galactic coordinates of
stars is important. Consequently, the best data are
the coordinates of stars from the Hipparcos catalog
(ESA 1997).

The model by Arenou et al. was used to calculate
the extinctions for 89470 Hipparcos stars with parallaxes
exceeding 0.0025 arcsec (i.e., located approximately
in the Gould Belt). Based on these data, we obtained
a solution to the system of equations (3) that is in
best agreement with the model by Arenou et al. It is
presented in the table as the HIP solution. Figure 4
shows how well the extinctions calculated from this
solution agree with those inferred from the model by
Arenou et al.: (a) for 89470 Hipparcos stars with
parallaxes exceeding 0.0025 arcsec and (b) for 111444 stars
with parallaxes exceeding 0.0005 arcsec.

The standard deviation of the differences between
the extinctions calculated using the two models is
designated in the table as $\sigma(A_{G}-A_{Arenou})$. It is $0.13^{m}$.
This confirms that the parabolas of the model by Arenou
et al. can be replaced by the suggested analytical
expression with two sine waves. In this case, the
orientation of the Gould Belt, the thickness of the
absorbing layers, the displacement of the Sun, and
the total extinction correspond to the expected ones.


\begin{table}
\caption[]{Solutions of the system of equations (3)}
\label{solu}
\[
\begin{tabular}{lcccccc}
\hline
\noalign{\smallskip}
                    & HIP & GCS & V86 & $OB_{hip}$ & $OB_{rpm}$ & $OB_{ph}$ \\
\hline
\noalign{\smallskip}
$\gamma$, deg               & $14\pm2$    & $17\pm2$    & $17\pm2$    & $17\pm2$    & $15\pm2$    & $19\pm2$    \\
$\lambda_{0}$, deg          & $-10\pm2$   & $-2\pm7$    & $-10\pm3$   & $-8\pm3$    & $-8\pm3$    & $-10\pm3$   \\
$Z_{A}$, pc                      & $70\pm20$   & $63\pm14$   & $100\pm30$  & $90\pm20$   & $60\pm20$   & $50\pm20$   \\
$\zeta_{A}$, pc                  & $60\pm20$   & $26\pm10$   & $30\pm30$   & $50\pm20$   & $60\pm20$   & $45\pm20$   \\
$Z_{0}$, pc                      & $15\pm3$    & $4\pm11$    & $-7\pm10$   & $16\pm6$    & $10\pm6$    & $0\pm10$    \\
$\zeta_{0}$, pc                  & $3\pm3$     & $0\pm7$     & $0\pm10$    & $6\pm3$     & $0\pm3$     & $0\pm10$    \\
$A_{0}$, mag pc$^{-1}$                 & $1.2\pm0.1$ & $0.8\pm0.1$ & $2.0\pm0.3$ & $2.2\pm0.1$ & $1.9\pm0.2$ & $2.1\pm0.2$ \\
$A_{1}$, mag pc$^{-1}$                 & $0.6\pm0.1$ & $0.3\pm0.1$ & $0.4\pm0.3$ & $0.6\pm0.2$ & $0.7\pm0.1$ & $0.9\pm0.2$ \\
$A_{2}$, deg                & $45\pm5$    & $39\pm7$    & $35\pm10$   & $37\pm3$    & $30\pm4$    & $32\pm5$    \\
$\Lambda_{0}$, mag pc$^{-1}$           & $1.0\pm0.1$ & $0.0\pm0.1$ & $0.2\pm0.2$   & $0.1\pm0.1$ & $0.5\pm0.2$ & $0.2\pm0.2$ \\
$\Lambda_{1}$, mag pc$^{-1}$           & $0.9\pm0.1$ & $0.4\pm0.1$ & $1.2\pm0.4$ & $1.0\pm0.1$ & $1.0\pm0.1$ & $0.8\pm0.2$ \\
$\Lambda_{2}$, deg          & $130\pm5$   & $129\pm8$   & $130\pm10$  & $131\pm5$   & $140\pm5$   & $133\pm5$   \\
$\sigma(A_{G}-A_{Arenou}), ^{m}$ & 0.13        &             &             &             &             &             \\
$\sigma(A_{Obs}-A_{Arenou}), ^{m}$ &             & 0.11        & 0.27        & 0.26        & 0.37        & 0.35        \\
$\sigma(A_{Obs}-A_{G}), ^{m}$      &             & 0.06        & 0.28        & 0.22        & 0.38        & 0.35        \\
\hline
\end{tabular}
\]
\end{table}


\section*{COMPARISON OF THE MODELS WITH OBSERVATIONS}

Having ascertained that the two extinction models
agree for a large number of Hipparcos stars, let
us compare the models with the extinction data for
real stars. Unfortunately, the extinction is low within
500 pc of the Sun, where the Gould Belt is located.
Therefore, it is determined with a low relative accuracy,
irrespective of the method of its determination.
Consequently, there are few accurate extinction estimates
for this region of space. In this paper, we
consider three catalogs of stars with fairly accurate
individual extinctions: for OB stars from the Hipparcos
and Tycho-2 catalogs (H\o g et al. 2000), the
extinctions were determined frommultiband photometry
by Gontcharov (2008a; hereafter OB stars); for F
and G dwarfs from the Geneva-Copenhagen survey
of the solar neighborhood, the extinctions were
estimated from Str\"omgren photometry (Nordstr\"om
et al. 2004; hereafter GCS); and for various stars,
the extinctions were derived from $UBV$ photometry
and spectral classification (Guarinos 1992; hereafter
the V86 catalog, according to the Strasbourg
database). We selected stars with Hipparcos parallaxes
$\pi>0.002$ arcsec from GCS and V86. For OB stars,
we calculated the distances from their Hipparcos parallaxes
as well as the photometric and photoastrometric
distances (from reduced proper motions). Applying
the extinction model to OB stars is also a
check of how accurate the extinctions and distances
calculated for them are.

Below, we will denote the observed extinction by
$A_{Obs}$, the extinction from the model by Arenou et
al. (1992) by $A_{Arenou}$, and the extinction from our
analytical model by $A_{G}$.

\subsection*{GCS}

The extinctions for GCS stars were determined
from multiband Str\"omgren photometry. The
solution of Eqs. (3) for 9996 of these stars with $\pi>0.002$ arcsec from Hipparcos is presented in the table as the
GCS solution. Here, the standard deviations of the
$A_{Obs}-A_{Arenou}$ and $A_{Obs}-A_{G}$ differences are denoted
by $\sigma(A_{Obs}-A_{Arenou})$ and $\sigma(A_{Obs}-A_{G})$, respectively.
We see that the suggested analytical model agrees
better with the data than the model by Arenou et al.
Unfortunately, the high accuracy of the extinctions
determined from Str\"omgren photometry is compensated
for by the closeness of the stars under consideration
and, accordingly, by the low value of the
extinction itself. In addition, the observed extinction
is systematically lower than that calculated from both
models: the mean observed extinction is $A_{V}=0.04^{m}$,
while the extinctions from the models by Arenou et al.
and the suggested model are $0.12^{m}$ and $0.17^{m}$, respectively.
This may be the result of selection in favor of
stars with lower extinctions in the GCS catalog.

\subsection*{V86}

We determined the extinctions for V86 stars,
which represent the entire variety of stellar types,
from $UBV$ photometry and spectral classification.
The solution of Eqs. (3) for 9319 V86 stars with $\pi>0.002$ arcsec from Hipparcos is presented in the table as the
V86 solution. Contrary to popular belief, the present day
spectral type-color index calibrations have an
accuracy of no better than $\pm0.3^{m}$ (or even $1^{m}\div2^{m}$)
due to the natural scatter of stellar characteristics
(Perryman 2009, p. 215). The values of $\sigma(A_{Obs}-A_{Arenou})$ and $\sigma(A_{Obs}-A_{G})$ given in the table confirm
the crucial role of the natural scatter of stellar characteristics.
The mean observed extinction is $A_{V}=0.28^{m}$, the extinction from the model by Arenou et al.
is $0.25^{m}$, and the extinction from the suggested model
is $0.38^{m}$.

\subsection*{OB stars}

In this study, we are interested not in
the spatial distribution of stars and other statistical
characteristics of the sample of OB stars from Goncharov
(2008a) but in the extinction, whose accuracy
depends primarily on the accuracy of the multiband
photometry used. Therefore, for this study, the stars
with photometry less accurate than $0.05^{m}$ at least in
one of the Tycho-2 and 2MASS bands under consideration
were excluded from our sample: $B_{T}$, $V_{T}$, $J$, $H$, $Ks$. This should provide an accuracy of determining
the reddening $E_{(B-V)}$ at a level of $0.1^{m}$ when
four photometric quantities are used for this purpose.
However, Gontcharov (2008a) calculated the
extinction coefficient $R=A_{V}/E_{(B-V)}$ not for each
star but for a region of space and, as Wegner (2003)
showed, the standard deviation (natural scatter) of the
coefficient $R$ for OB stars in the same region of space
is typically more than $0.3^{m}$. Consequently, this is the
expected accuracy of determining the extinctions for
the OB stars considered here.

Instead of the relation $R=2.8+0.18\sin(l+115^{\circ})$ adopted by Gontcharov (2008a), here we use
an apparently more plausible relation, $R=2.65+0.2\sin(l+75^{\circ})$. Like the previous one, it was derived
by extrapolating the extinction law from the
$(B_{T}-V_{T})$ -- $(V_{T}-Ks)$ relation based on Tycho-2 and
2MASS photometry and is valid as an approximation
at low Galactic latitudes. Obviously, a separate
detailed study of the variations in $R$ will be necessary.

To improve the accuracy, the stars with known
spectral classification from the TST catalog (Wright
et al. 2003) that did not belong to the O, B, and A0
types (the latter were retained, because they could
actually belong to the B type due to the classification
errors) were excluded from our sample. As a result of
our analysis of the dependence of extinction on the effective
wavelength, we adopted the following relation
between the extinctions in the Johnson V band and
the Tycho-2 $V_{T}$ band in all our calculations: $A_{V}=0.884A_{V_{T}}$. We rejected the stars with an observed
extinction exceeding $1.1/(0.25+|\sin(b)|)$; as a rule,
these are peculiar late-type stars.

15670 stars remained in the final sample, instead of 37485 stars in the initial sample.

In Fig. 5, the observed extinction is plotted against
the photometric distance forGalactic latitudes $-5^{\circ}<b<+5^{\circ}$ in the first (a), fourth (b), and sixth (c)
Galactic octants, which differ significantly by the
pattern of extinction. We see a large natural scatter
of individual extinctions in the first octant (more
than $0.5^{m}$), a much ``quieter'' picture in the fourth
octant, and a low extinction and a standard deviation
of less than $0.3^{m}$ in the sixth octant. The mean
deviation of the individual stellar extinction from the
analytical model also changes accordingly.

In Fig. 6, the extinction is plotted against the
Galactic longitude. This dependence corresponds to
the data of the model by Arenou et al. (1992) shown
in Fig. 2a.

For 2472 Hipparcos OB stars with parallaxes exceeding
$0.0025$ arcsec, we obtained a solution presented in
the table as the $OB_{hip}$ solution.

Figure 7 shows the correspondence between the
various extinctions for these 2472 OB stars: (a) $A_{Obs}$
and $A_{Arenou}$; (b) $A_{Obs}$ and $A_{G}$ calculated from the
analytical model using photoastrometric distances;
(c) $A_{Arenou}$ and $A_{G}$ calculated from the analytical
model using Hipparcos parallaxes; and (d) $A_{Obs}$ and
$A_{G}$ calculated from the analytical model using Hipparcos
parallaxes. On the whole, we see good agreement.

The group of stars deviates from the bisector in
Fig. 7c due to the asymmetry of the model by Arenou
et al. (1992) in the Gould Belt, which is also seen in
Fig. 2a: in the region with $l\approx180^{\circ}$, $b\approx-15^{\circ}$, the
extinction is $A_{V}\approx2^{m}$, i.e., approximately twice that
in the symmetric region with $l\approx0^{\circ}$, $b\approx+15^{\circ}$,
where $A_{V}\approx1^{m}$. Our new analytical model gives approximately
equal extinctions for these regions.

The group of stars deviating from the bisector in
Fig. 7d consists mostly of supergiants and bright
OB giants. The enhanced extinction in this case may
be explained by the existence of an additional, noninterstellar
extinction, say, in the envelopes of these
stars.

For 7615 stars with photoastrometric distances
less than 400 pc, we obtained a solution presented in
the table as the $OB_{rpm}$ solution. For 3054 stars with
photometric distances less than 400 pc, we obtained
a solution presented in the table as the $OB_{ph}$ solution.

\section*{DISCUSSION}

The following conclusions can be reached by comparing
the solutions in the table.

(1) The differences in the solutions and the low
accuracy of determining some of the unknowns confirm
that the analytical models of extinction within the
nearest kiloparsec are difficult to test, because there
are no accurate extinction determinations for a large
number of stars.

(2) The acceptable accuracy of the $OB_{rpm}$ and
$OB_{ph}$ solutions as well as their agreement with the
remaining solutions suggest that the photoastrometric
and photometric distances can be used for an approximate
calculation of the extinctions of stars based
on the suggested analytical model if more accurate
distances are unavailable.

(3)Comparison of $\sigma(A_{Obs}-A_{Arenou})$ and $\sigma(A_{Obs}-A_{G})$ shows that the suggested model agrees with the
observational data no more poorly than the model by
Arenou et al. (1992).

(4) As expected, the two models under consideration
are closer to each other than to the observational
data due to the natural scatter of extinctions for individual
stars.

(5) The values found for 10 unknowns (except $A_{0}$ and $\Lambda_{0}$) agree in all solutions.

(6) $A_{0}$ and $\Lambda_{0}$ vary from one solution to another,
but their sum, along with the total calculated extinction,
are approximately the same in all of the solutions
except the GCS one. Only the total constant
extinction rather than the constant extinction in each
layer is reliably determined probably because of the
low inclination of the Gould Belt to the Galactic
plane and, accordingly, the narrow range of longitudes
where the absorbing layers are separated. Since
the extinction for GCS stars is low, we could not
correctly estimate $A_{0}$ and $\Lambda_{0}$. It is clear from physical
considerations that $A_{0}$ and $\Lambda_{0}$ must be approximately
equal. Therefore, we finally adopted $A_{0}=1.2^{m}$, $\Lambda_{0}=1.1^{m}$.

The final formulas to calculate the extinction from
the suggested model with the best values of the unknowns
found here are
$$ (1.2+0.6\sin(l+35^{\circ}))r(1-e^{-|Z-0.01|/0.07})0.07/|Z-0.01| $$
and
$$ (1.1+0.9\sin(2\lambda+135^{\circ}))r(1-e^{-|\zeta|/0.05})0.05/|\zeta|, $$
where the longitudes $l$ and $\lambda$ are given in degrees;
the distances $r$, $Z$, and $\zeta$ are given in kpc, and the
resulting extinction is given in magnitudes. Figure 2b
shows the dependence of extinction $A_V$ on the Galactic
longitude calculated from the suggested model for
a distance of 500 pc and various Galactic latitudes.
The vertical bars indicate a relative accuracy of 40\%.
In Figs. 2a and 2b, we see good agreement between
the models under consideration.

The extinction in the Gould Belt must hide the
more distant stars. Indeed, analysis of the number
of stars and the extinction for the OB stars under
consideration with photometric distances from 400 to
800 pc shows that the number of stars is considerably
larger in the southern (relative to the Sun and not
relative to the Galactic equator!) hemisphere: 3841
versus 3415, while the extinction is higher in the
northern hemisphere: $0.76^{m}$ versus $0.66^{m}$. Figure 8
shows smoothed contour maps for the stars under
consideration within 400--800 pc and $-36^{\circ}<b<+36^{\circ}$: (a) the extinction and (b) the number of stars.
We see that the regions with a large number of stars
have a comparatively low extinction and vice versa,
i.e., the stars in the high-extinction regions are hidden.
The high-extinction regions are located predominantly
above the Galactic equator toward $l\approx0^{\circ}$ and
below it toward $l\approx180^{\circ}$. This reflects the orientation
of the Gould Belt. The stars remaining in the transparency
windows do not show any distribution along
the Gould Belt in Fig. 8b. As a result, contrary to
the universally accepted approach, it is hard to judge
the size and orientation of the Gould Belt from the
apparent spatial distribution of its constituent stars.
The Belt may have larger sizes than the commonly
assumed ones.

The height of the Sun above the Galactic equator
is usually estimated by counting the stars with
positive and negative latitudes (Veltz et al. 2008).
However, our study shows that the layer with the
highest extinction may be several parsecs above the
Sun, which, in turn, is approximately 13 pc above
the Galactic plane (Gontcharov 2008b). At such a
location, the northern stars, especially the distant
ones, are hidden by extinction to a greater extent than
the southern stars, and the star counts should give a
height of the Sun above the Galactic plane larger than
the true value, as, for example, 20 pc in Veltz et al.
(2008). Obviously, when the height of the Sun above
the Galactic plane is determined, the noncoincidence
of this plane with the layer of the highest extinction
should be taken into account.

\section*{CONCLUSIONS}

This study has shown that there is absorbing matter
both in the Galactic plane and in the plane of
the Gould Belt. For this reason, the Gould Belt contributes
significantly to the extinction for stars at
moderate Galactic latitudes, especially toward the
Galactic center (at positive latitudes) and anticenter
(at negative latitudes). The 3D model by Arenou et
al. (1992) that represents the extinction as a parabola
depending on the distance with various coefficients
for 199 regions of the sky can be replaced by the
sum of two sine waves that describe the extinction
in the Galactic plane and in the plane of the Gould
Belt. The absorbing layers were found to intersect at
an angle of about $17^{\circ}$. This angle may be considered
as the inclination of the Gould Belt to the Galactic
equator. The angle between the line of intersection of
the Gould Belt with the Galactic plane and the Y axis
is about $-10^{\circ}$. The Sun is probably several parsecs
below the axial plane of the equatorial absorbing layer
but above the Galactic plane.

The suggested analytical 3D extinction model
agrees with the observed extinction for stars from
the three catalogs considered and gives an extinction
estimate for any star within 500 pc of the Sun (and
possibly farther) based on its Galactic coordinates.

This paper opens a series of studies of the interstellar
extinction in our Galaxy. In the future, we are
going to construct an extinction model for much of
the Galaxy using 2MASS photometry, which, as is
shown Fig. 1, allows the reddening of stars in different
directions and at different distances to be estimated.

\section*{ACKNOWLEDGMENTS}

In this study, we used various results from the Hipparcos
project and data from 2MASS (Two Micron All Sky Survey), a joint project of the Massachusetts
University and the IR Data Reduction and Analysis Center of the California Institute of Technology
financed by NASA and the National Science Foundation.
We also used resources from the Strasbourg Data Center (France) (http://cdsweb.u-strasbg.fr/).
This study was supported in part by the Russian Foundation for Basic Research (http://www.rfbr.ru)
(project no. 08-02-00400) and in part by the ``Origin and Evolution of Stars and Galaxies'' Program of the
Presidium of the Russian Academy of Sciences.

\newpage

\begin{figure}
\includegraphics{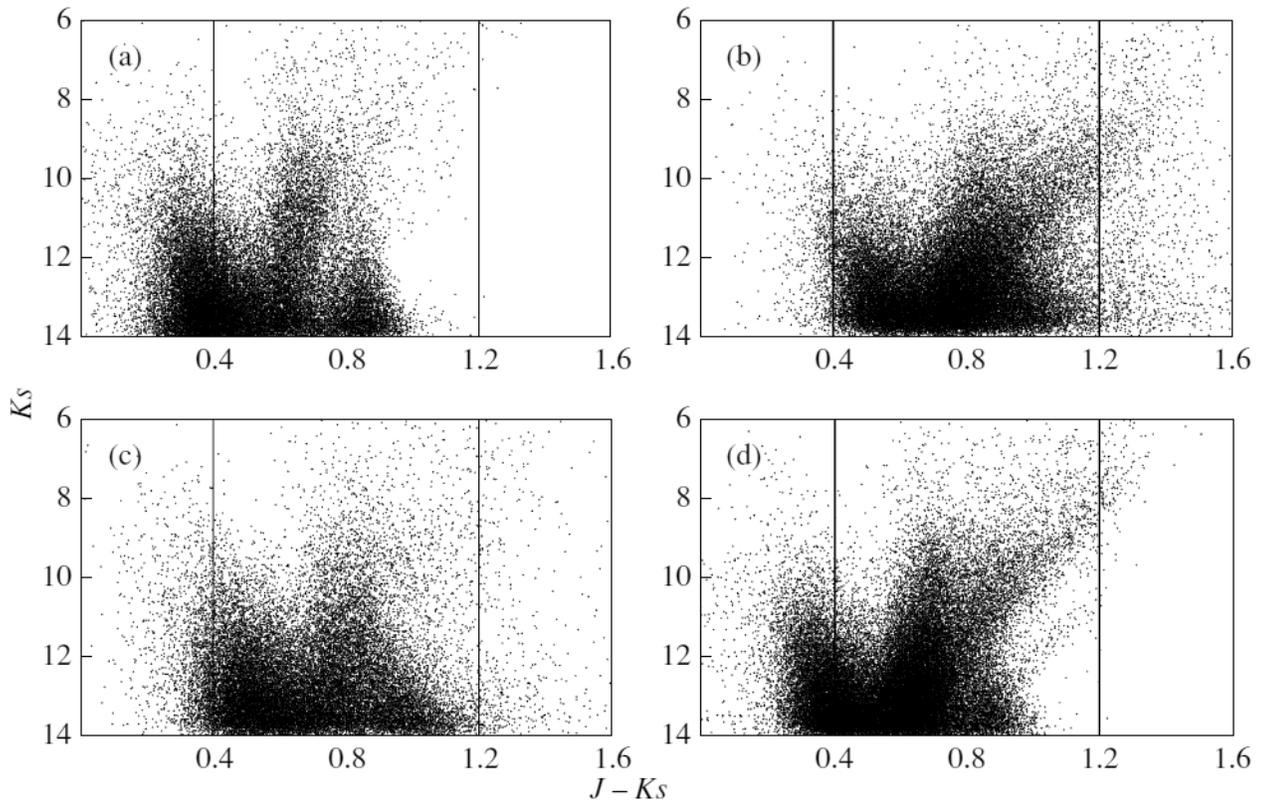}
\caption{$(J-Ks)$ -- $Ks$ diagram for 2MASS stars with accurate photometry:
(a) near $b=+15^{\circ}$, $l=180^{\circ}$;
(b) near $b=+15^{\circ}$, $l=0^{\circ}$;
(c) near $b=-15^{\circ}$, $l=180^{\circ}$;
(d) near $b=-15^{\circ}$, $l=0^{\circ}$.
}
\label{jkk}
\end{figure}

\begin{figure}
\includegraphics{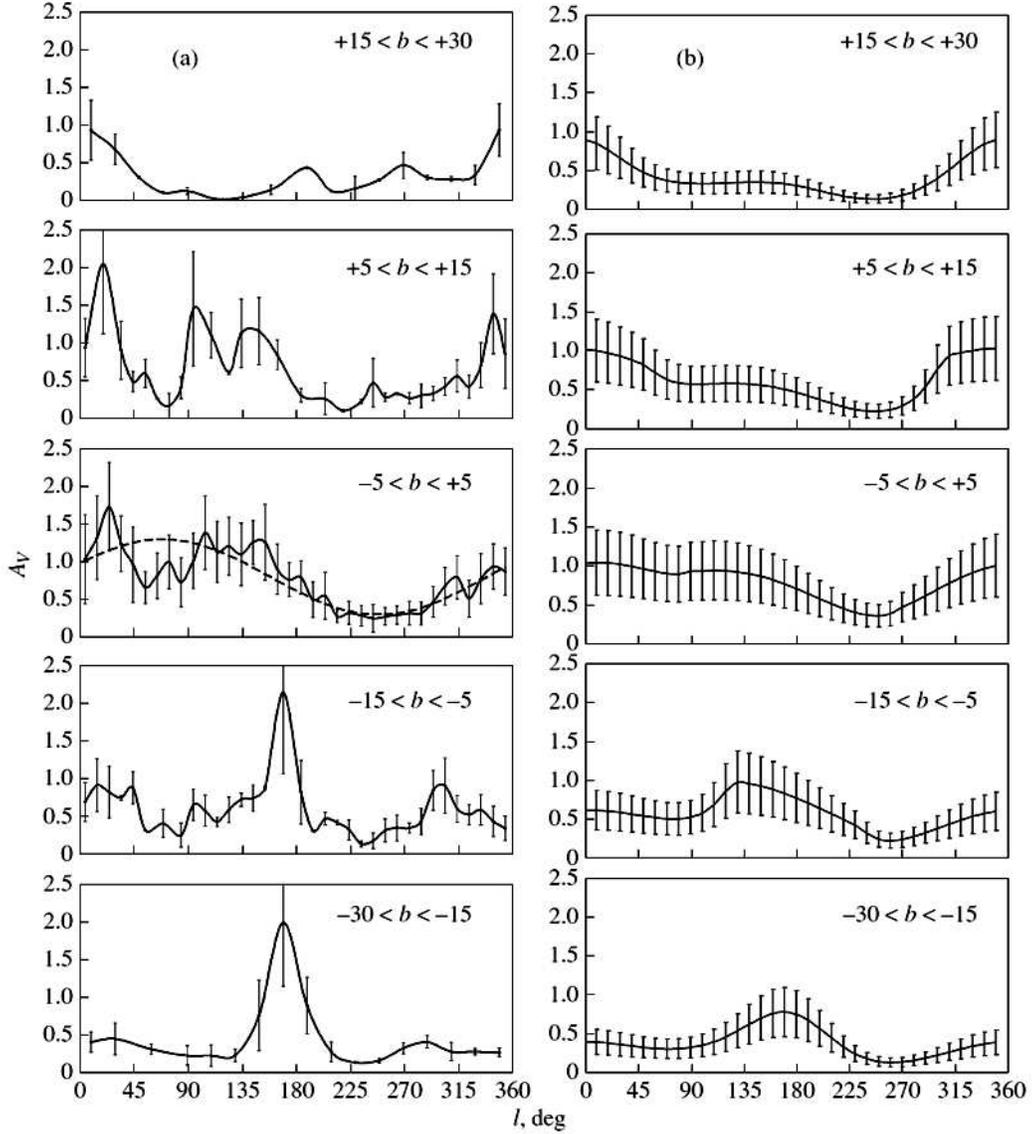}
\caption{Extinction $A_V$ versus Galactic longitude at a distance of 500 pc from the Sun for various Galactic latitudes:
(a) according to the model by Arenou et al. (1992);
(b) according to the model suggested here. The dashes indicate the dependence
$0.8+0.5\sin(l+20^{\circ})$. The vertical bars indicate the accuracy of the model by Arenou et al. (1992) and
a relative accuracy of 40\% for the suggested model.
}
\label{model}
\end{figure}

\begin{figure}
\includegraphics{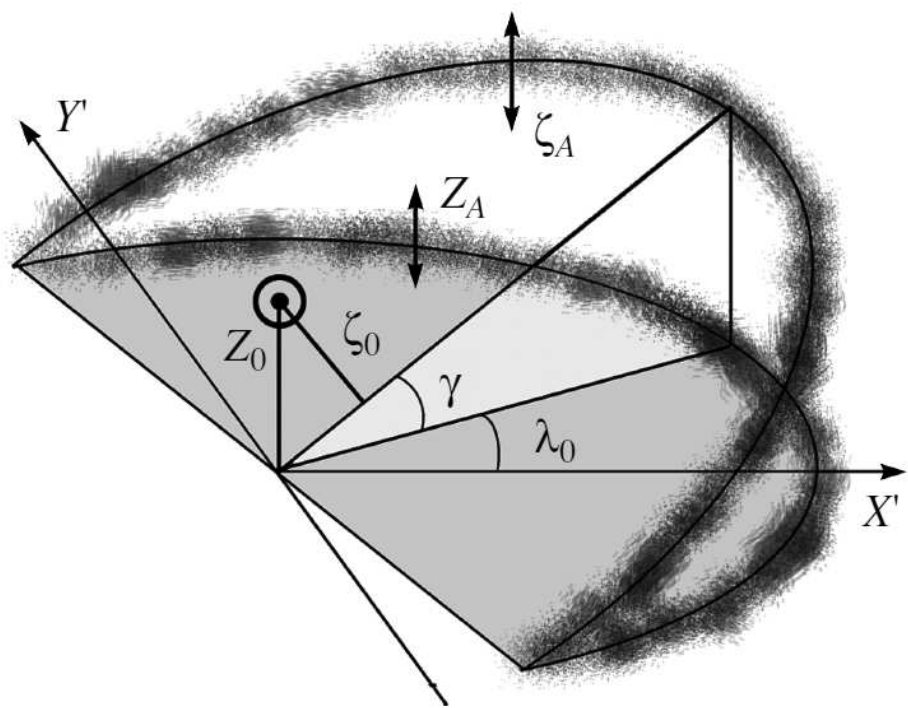}
\caption{Relative positions of two absorbing layers.
}
\label{intro}
\end{figure}

\begin{figure}
\includegraphics{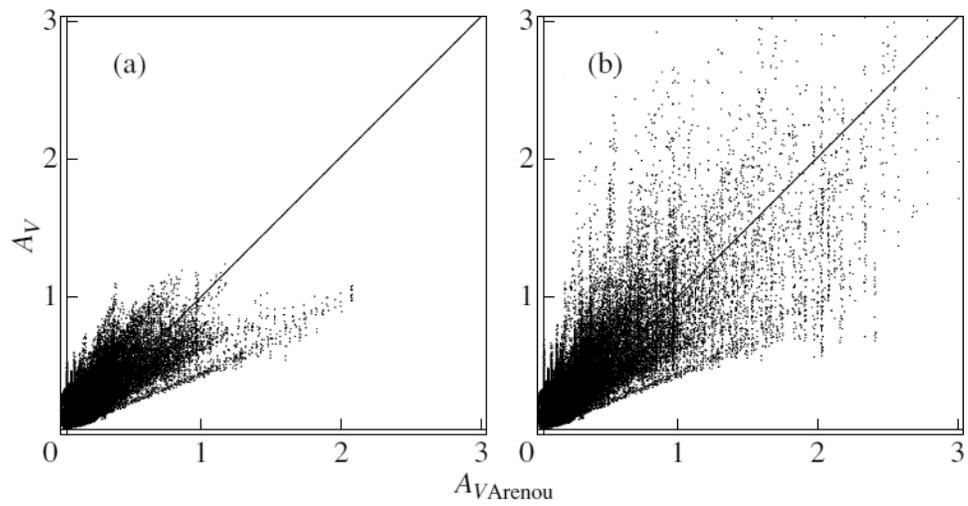}
\caption{Correspondence between the extinctions calculated here and those inferred from the model by Arenou et al.
(1992) for 89470 Hipparcos stars with parallaxes exceeding 0.0025 arcsec (a)
and for 111444 stars with parallaxes exceeding 0.0005 arcsec (b).
}
\label{amodel}
\end{figure}

\begin{figure}
\includegraphics{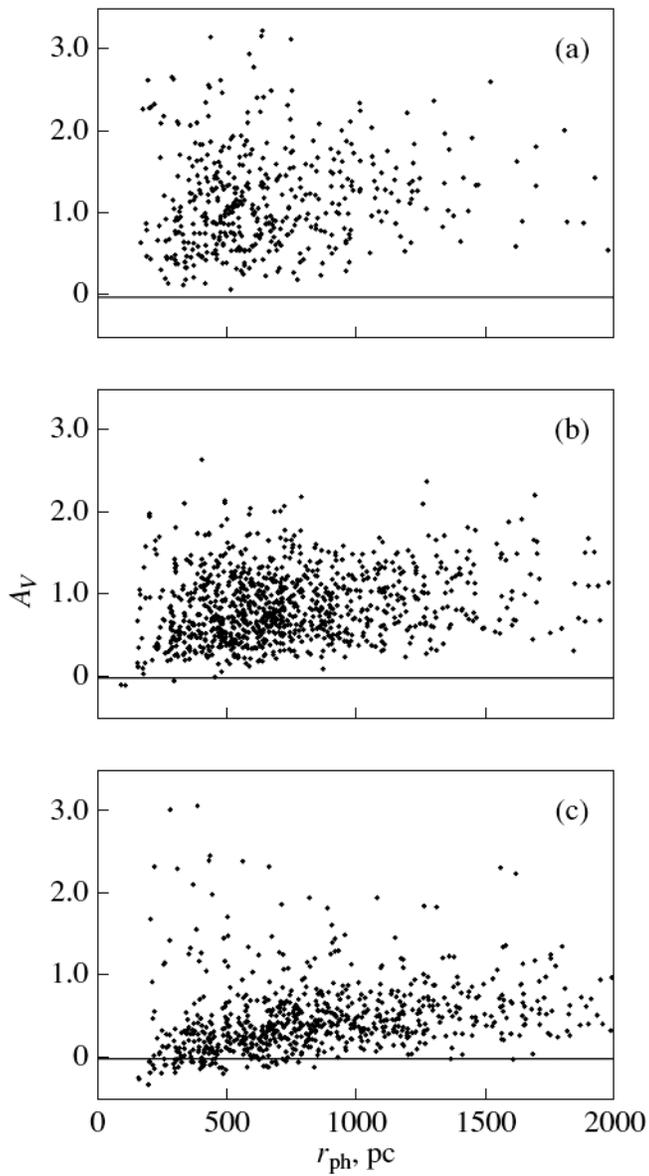}
\caption{Extinction versus photometric distance for the
OB stars under consideration in the region $-5^{\circ}<b<+5^{\circ}$ for (a) $0^{\circ}<l<45^{\circ}$,
(b) $135^{\circ}<l<180^{\circ}$, and (c)
$225^{\circ}<l<270^{\circ}$.
}
\label{rav}
\end{figure}

\begin{figure}
\includegraphics{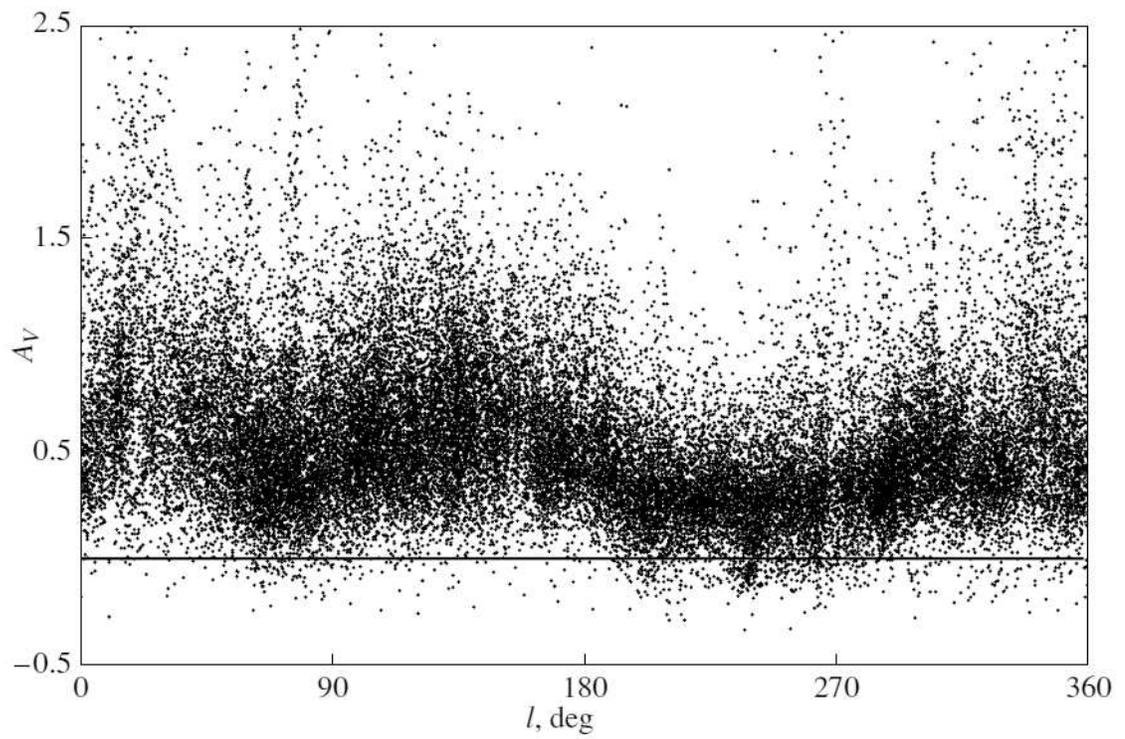}
\caption{Extinction versus Galactic longitude for the OB stars under consideration.
}
\label{av}
\end{figure}

\begin{figure}
\includegraphics{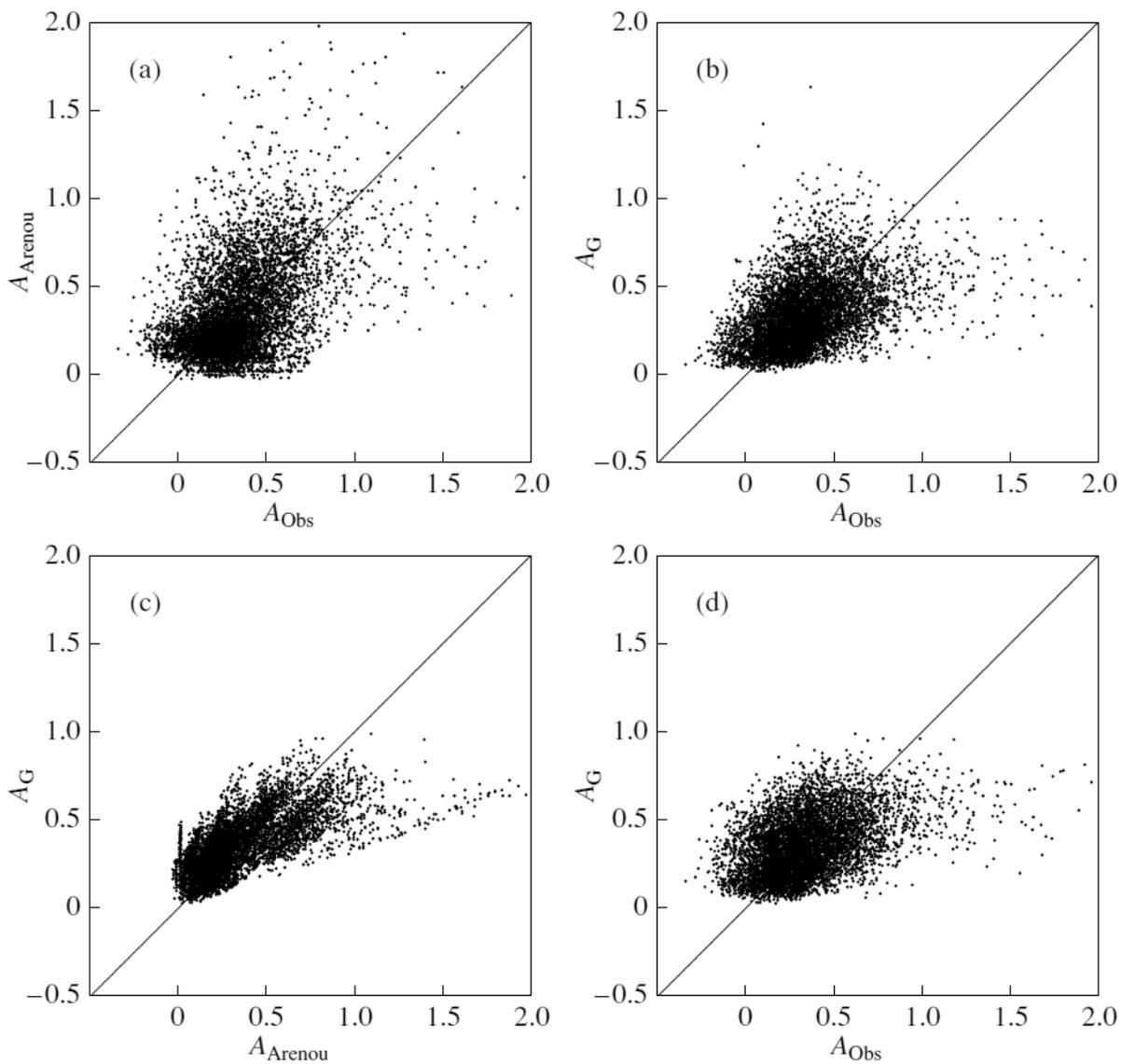}
\caption{Correspondence between the various extinctions for 2472 Hipparcos stars:
(a) the observed extinctions and those from Arenou et al. (1992);
(b) the observed extinctions and those calculated using photoastrometric distances;
(c) the extinctions from Arenou et al. (1992) and those calculated using astrometric distances;
(d) the observed extinctions and those calculated using astrometric distances.
}
\label{ob1}
\end{figure}

\begin{figure}
\includegraphics{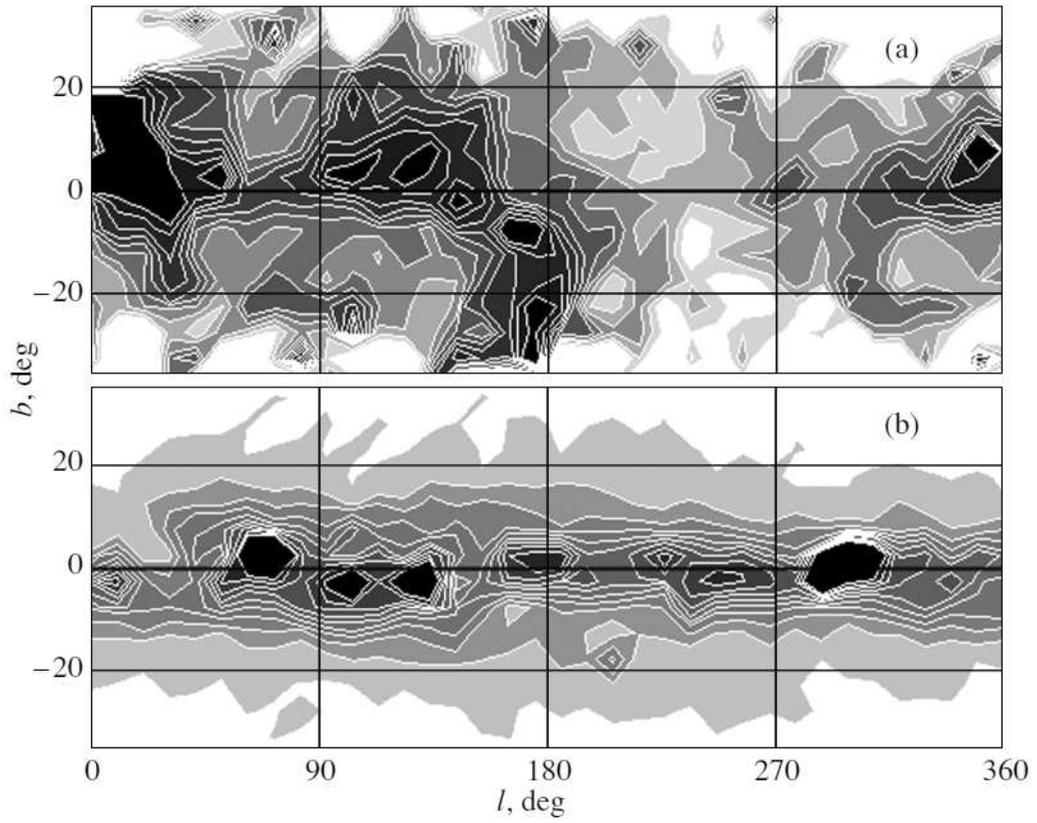}
\caption{Smoothed contour maps for the OB stars under consideration within 400--800 pc and
$-36^{\circ}<b<+36^{\circ}$: (a) the extinction and (b) the number of stars.
}
\label{an}
\end{figure}

\end{document}